\documentclass[preprint,prd,nofootinbib]{revtex4-1}
\usepackage{epsfig,graphicx,amssymb}
 \def\lsim{\mathrel{\vcenter{\hbox{$<$}\nointerlineskip\hbox{$\sim$}}}}
\def\gsim{\mathrel{\vcenter{\hbox{$>$}\nointerlineskip\hbox{$\sim$}}}}
\begin{document}
\begin{flushright}
	UMD-PP-012-001\\
	January 2012
\end{flushright}
\title{Multi-Lepton Collider Signatures of Heavy Dirac and Majorana Neutrinos}
\author{Chien-Yi Chen$^{1,}$\footnote{\tt chienyic@andrew.cmu.edu} and P. S. Bhupal Dev$^{2,}$\footnote{\tt bhupal@umd.edu}}
\affiliation{\vspace{0.2cm}
$^1$Department of Physics, Carnegie Mellon University, Pittsburgh, PA 15213, USA}
\affiliation{$^2$Maryland Center for Fundamental Physics and 
Department of Physics, University of Maryland, College Park, MD 20742, USA}

\begin{abstract}
We discuss the possibility of observing multi-lepton signals at the Large 
Hadron Collider (LHC) from the 
production and decay of heavy Standard Model (SM) singlet neutrinos added in 
extensions of SM to explain the observed light neutrino 
 masses by seesaw mechanism. 
In particular, we analyze two `smoking gun' signals depending on the Dirac or 
Majorana nature of the heavy neutrino: (i) for Majorana case, the  
same-sign di-lepton signal which can be used as a probe of lepton-number 
violation, and (ii) for Dirac case, the tri-lepton signal which conserves lepton number 
but may violate lepton flavor. Within a minimal Left-Right symmetric framework 
in which these additional neutrino states arise naturally, we find that in both cases, the 
signals can be identified with virtually no background beyond a TeV, and  
the heavy gauge boson $W_R$ can be discovered in this process. This analysis 
also provides a direct way to probe the nature of seesaw physics 
involving the SM singlets at TeV scale, and in particular, to distinguish type-I 
seesaw with purely Majorana heavy neutrinos from inverse seesaw with pseudo-Dirac 
counterparts. 
\end{abstract}

\maketitle
\section{Introduction}
\label{intro} 
One of the major evidences for the existence of new physics beyond the 
Standard Model (SM) is the discovery of non-zero neutrino mass from the  
observation of neutrino flavor oscillation phenomenon (for a recent update 
on the global neutrino data analysis, see Ref.~\cite{global}). 
In the SM, the left-handed (LH) 
neutrinos are massless mainly due to the absence of their right-handed (RH) 
counterparts (hence no Dirac mass) as well as the conservation of a global
$B-L$ symmetry (hence no Majorana mass). Therefore, in order to generate 
non-zero neutrino masses, one must extend the SM sector by either adding 
three RH neutrinos  
(one per family) or by introducing $(B-L)$-breaking fields or both~
\cite{review}. 
If we just add RH neutrinos ($N$) while keeping the $B-L$ symmetry unbroken, then 
the observed smallness of LH-neutrino masses require that the new Yukawa
couplings ($h_\nu$) 
involving the interaction of the $N$'s with the LH-doublet ($L$) 
given by $h_\nu\bar{L}HN$ (where $H$ is the SM Higgs doublet) 
must be extremely small, i.e. 
$h_\nu\lsim 10^{-12}$ for sub-eV LH neutrino mass. 
In the absence of any obvious compelling arguments
for such a tiny Yukawa coupling, the alternative path of generating non-zero 
neutrino masses by breaking $B-L$ symmetry seems more natural.  

The simplest way to parameterize the $B-L$ breaking effects in SM extensions 
is through Weinberg's dimension-5 operator~\cite{weinberg} 
\begin{eqnarray}
	{\cal L}_{\rm eff} = \lambda_{ij}\frac{L_i L_j HH}{M}~ ~ ~ ~ (i,j=e,\mu,\tau)
	\label{eq:1}
\end{eqnarray}
added to the SM Lagrangian, where $M$ is the scale of new physics. After 
electroweak symmetry breaking (EWSB) due to the Higgs vacuum expectation value (vev) 
$\langle H \rangle\equiv v_{\rm wk}$, 
this operator leads to a non-zero 
neutrino mass of the form $m_\nu = \lambda v^2_{\rm wk}/M$,  
and hence, $M/\lambda \gsim 10^{14}$ GeV for sub-eV neutrino mass. Thus, 
the new physics scale $M$ depends on the effective Yukawa coupling $\lambda$ 
(which is model-dependent), 
and can be in the TeV range to be directly accessible at colliders provided 
$\lambda$ is very small.

There are both tree- and loop-level realizations of the dimension-5 
operator given by 
Eq.~(\ref{eq:1}) to generate non-zero neutrino masses~\cite{ma1}. The 
tree-level realization is the well-known 
seesaw mechanism in which the heavy particles associated with the new physics,  
after being integrated out, lead to the effective operator in  
Eq.(\ref{eq:1}). The simplest such model is the type I seesaw~\cite{type1} in 
which the heavy particles are SM singlet Majorana fermions, usually known as 
the 
RH neutrinos ($N$), which couple to LH-doublets through Dirac Yukawa: 
\begin{eqnarray}
	{\cal L}_Y = (h_\nu \bar{L} H N + {\rm h.c.}) + NM_N N,
\end{eqnarray}
and $M_N$ is the Majorana mass of $N$. After EWSB, this leads to the neutrino mass matrix 
of the form 
\begin{eqnarray}
	{\cal M}_\nu = \left(\begin{array}{cc}
		0 & M_D\\
		M_D^T & M_N
	\end{array}\right),
	\label{eq:type1}
\end{eqnarray}
where $M_D=v_{\rm wk}h_\nu$. The light mass eigenvalues are given by 
\begin{eqnarray}
	m_\nu = -v_{\rm wk}^2 h_\nu M_N^{-1}h_\nu^T.
\label{eq:3}
\end{eqnarray}
It is clear from Eq.~(\ref{eq:3}) that for TeV scale $M_N$, 
the Dirac Yukawa $h_\nu\lsim 10^{-6}$, 
unless there are cancellations to get small neutrino masses from large Dirac 
masses using symmetries~\cite{kersten}. The heavy RH neutrinos, 
being SM singlets, can be produced at colliders only via $\nu-N$ mixing after 
virtual $W (Z)$'s produced in parton collision decay to $\ell (\bar{\nu})+ 
\nu$. Once produced, the $N$'s decay equally likely to both charged leptons 
and anti-leptons (due to their Majorana nature), thus giving the distinct 
collider signature of like-sign di-lepton final states\footnote{This is a
collider analogue of neutrinoless double beta decay to probe the lepton number 
violation.}. However, the mixing 
in type-I seesaw is typically given by $\theta_{\nu N} 
\sim \sqrt{m_\nu M_N^{-1}} \lsim 
10^{-6}$ (again barring cancellations), and hence, the production of the $N$'s 
is highly suppressed. A detailed collider simulation shows that 
the minimal type I seesaw can be tested at colliders only if 
$\theta_{\nu N}$ is 
large ($\gsim 10^{-2}$) or $M_N$ is small (up to a few hundred GeV)~
\cite{mtype1col,mtype1colb,type13col}.

A second way to write the Weinberg operator in Eq.~(\ref{eq:1}) is  
$(L^T\vec{\tau}L)\cdot(H^T\vec{\tau}H)/M$ where $\tau^i$'s are the usual Pauli 
matrices. This can be implemented by adding an
$SU(2)_L$ bosonic triplet $\vec{\Delta} \equiv (\Delta^{++},\Delta^+,\Delta^0)$
coupled to SM leptons through Majorana type 
couplings. This is known as the type II seesaw mechanism~\cite{type2}. The 
$\Delta$'s, being SM non-singlets, can couple directly to the SM gauge bosons 
($W,Z$ and $\gamma$), and can be easily produced at colliders if their 
masses are in the sub-TeV to TeV range. The presence of doubly and singly 
charged scalars in the triplet lead to a 
very rich collider phenomenology of such models~\cite{type2col} which 
can be easily explored at the LHC. 

Yet another way to write the effective Weinberg operator in 
Eq.~(\ref{eq:1}) is $(L^T\vec{\tau}H)^2/M$ which can be implemented 
by adding an $SU(2)_L$ fermionic triplet ($\vec{\Sigma}$) coupled 
to leptons through Dirac Yukawas, just like the singlet ones in type I. This 
is known as the type III seesaw~\cite{type3}, and has very similar collider 
signatures as in type I case, except for the fact that the triplet fermions 
in this case couple directly to the SM $W$-boson, which makes them easier to 
search for at colliders up to about a TeV mass~\cite{type13col,type3col}. 

A completely different realization of the seesaw mechanism is the so-called
inverse seesaw mechanism~\cite{inverse}, where instead of one set of
SM singlet Majorana fermions, one introduces two sets of them: 
  $N$ (Dirac) and $S$ (Majorana). The resulting Lagrangian is given by 
  \begin{eqnarray}
	  {\cal L}_Y = (h_\nu \bar L HN+NM_NS+{\rm h.c.}) + S\mu S
	  \label{eq:4p}
  \end{eqnarray}
Due to the existence of the second set of singlet fermions (and perhaps
additional gauge symmetries), the neutrino mass
formula in these models has the form 
\begin{eqnarray}
	{\cal M}_\nu = \left(\begin{array}{ccc}
		0 & M_D & 0\\
M_D^T & 0 & M_N\\
0 & M_N^T & \mu
\end{array}\right)
\label{eq:invseesaw}
\end{eqnarray}
In the limit $\mu \ll v_{\rm wk} \lsim M_N$, the lightest mass eigenvalues are given by
  \begin{eqnarray}
	  m_\nu \simeq v^2_{\rm wk} h_\nu M_N^{-1}\mu\left(M_N^{-1}\right)^T h_{\nu}^T\equiv F\mu F^T
\label{eq:4}
\end{eqnarray}
where $\mu$ breaks the lepton number. Because of the presence of this new
mass scale in the theory which is directly proportional to the light 
neutrino mass, the seesaw scale $M_N$ can be naturally very low (within 
the range of colliders) even for ``large'' Dirac Yukawa couplings. This 
also allows for a large mixing $\theta_{\nu N}\simeq v_{\rm wk} h_{\nu}M_N^{-1}$, 
and makes the collider tests of this possibility much more feasible. 
However, due to the pseudo-Dirac nature of the RH neutrinos, the 
``smoking gun'' signal for type I seesaw, namely the lepton number violating 
same-sign di-lepton signal~\cite{mtype1colb} is 
absent in this case. Instead,  
the lepton flavor violating tri-lepton signal~\cite{trilepton1,trilepton2}
can be used to test these models. In this paper, we will mainly focus 
on these SM singlet RH neutrinos 
and present a detailed collider study of the 
multi-lepton final states in order 
to distinguish the heavy Dirac neutrinos from their Majorana counterparts at the LHC
\footnote{For a discussion on collider signals in other seesaw models, 
see Refs.~\cite{trilepton1,nath}.}.

Since the testability of seesaw is intimately related to the
magnitude of the seesaw scale and the couplings of the new 
heavy particles with the SM particles, a key  question of interest 
is whether there could be any theoretical guidelines for this new physics. 
A well-known example that explains the seesaw scale as a result of gauge 
symmetry breaking is the Left-Right (LR) Symmetric Theory based on the 
gauge group $SU(2)_L\times SU(2)_R\times U(1)_{B-L}$~\cite{LR}. 
Apart from restoring the parity symmetry at high energy, this theory 
provides a natural explanation of the seesaw scale as connected to the 
$SU(2)_R\times U(1)_{B-L}$-breaking scale. Also, the smallness of the 
neutrino mass is connected to the extent to which the RH-current is 
suppressed at low energy. Thus, the LR-symmetry provides a 
well-defined theory of neutrino masses~\cite{book} and can be used 
as a guide to study seesaw physics at the LHC~\cite{senja10}. Moreover, it provides a 
very attractive low-energy realization of a Grand Unified Theory (GUT) such as  
$SO(10)$, which is arguably the simplest GUT scenario for 
seesaw mechanism~\cite{review} as it automatically predicts the existence 
of RH neutrinos (along with the SM fermions). The $SO(10)$ embedding 
of TeV-scale LR models have been discussed in literature for both 
type I~\cite{type1gut} and inverse seesaw~\cite{invgut}. 
Also, in case of inverse seesaw, as pointed out in Ref.~\cite{adcm}, 
the LR gauge symmetry is essential to stabilize the form of the neutrino 
mass matrix given by Eq.~(\ref{eq:invseesaw}). Therefore, in this paper, we work within the 
framework of the minimal LR-symmetric theory at TeV-scale.

This paper is organized as follows: in Sec. II we briefly summarize the main 
features of the minimal LR model, including the mixing between light and heavy neutrinos as well 
as gauge bosons, relevant for our analysis; in Sec. III, we discuss the production 
and decay of a heavy SM singlet neutrino at the LHC; in Sec. IV, we perform a detailed collider 
simulation of the multi-lepton events; in Sec. V, we summarize our results.  

\section{The minimal Left-Right model}
In this section, we review the minimal LR model based on the gauge group 
$SU(2)_L\times SU(2)_R\times U(1)_{B-L}$~\cite{LR} and discuss the mixing between 
the light and heavy neutrinos as well as gauge bosons. We also set 
the notations for the following sections. 

In the LR model, the quarks and leptons are assigned to the following irreducible 
representations of the gauge group $SU(2)_L\times SU(2)_R\times U(1)_{B-L}$:
\begin{eqnarray}
	Q_L = \left(\begin{array}{c}
		u_L \\ d_L
	\end{array}
	\right) : (2,1,1/3), ~ ~ 
	Q_R =  \left(\begin{array}{c}
		u_R \\ d_R
	\end{array}
	\right) : (1,2,1/3),\nonumber\\
L_L = \left(\begin{array}{c}
		\nu_e \\ e_L
	\end{array}
	\right) : (2,1,-1), ~ ~ 
	L_R =  \left(\begin{array}{c}
		N_e \\ e_R
	\end{array}
	\right) : (1,2,-1)\nonumber
\end{eqnarray}
and similarly for second and third generations. The minimal Higgs sector 
consists of a bi-doublet $\Phi :(2,2,0)$ and two triplets 
$\Delta_L :(3,1,2)$ and $\Delta_R :(1,3,2)$. After the spontaneous 
breaking of the $SU(2)_L\times SU(2)_R\times U(1)_{B-L}$ symmetry 
to $U(1)_Q$ by the vev $v_{L,R}$ and 
$\kappa,\kappa'$ of the Higgs fields $\Delta^0_{L,R}$ and $\Phi$ respectively, 
the phenomenological requirement $v_L\ll \kappa,\kappa' \ll v_R$ 
ensures the suppression of the RH-currents and the smallness of the neutrino mass. 
Also, the LR symmetry $\psi_L \leftrightarrow \psi_R$ for fermions and 
$\Delta_L\leftrightarrow \Delta_R, \Phi \leftrightarrow \Phi^\dagger$ for the Higgs 
fields leads to the relations $g_L=g_R=e/\sin\theta_W$ 
and $g' = e/\sqrt{\cos{2\theta_W}}$ for the coupling strengths of the gauge 
bosons $W_{L,R}$ and $Z'$ corresponding to the $SU(2)_{L,R}$ and $U(1)_{B-L}$ 
gauge symmetries respectively (where $\theta_W$ is the Weinberg angle and $e$ is 
the electric charge of proton). 
\subsection{Mixing in the Gauge Sector}
The charged gauge bosons $W_{L,R}^\pm$ in the weak eigenstate mix in the 
mass eigenstates $W,W'$:
\begin{eqnarray}
	W &=& \cos{\zeta_W}W_L+\sin{\zeta_W}W_R,\nonumber\\
	W' &=& -\sin{\zeta_W}W_L+\cos{\zeta_W}W_R,
\end{eqnarray}
where $\tan{2\zeta_W} = 2\kappa\kappa'/(v_R^2-v_L^2)$. The current 
bound on the mixing angle is as low as $\zeta_W < 0.013$~\cite{zetaw,zetawp}; 
hence for our purposes, we can safely assume the mass eigenstates 
as the weak eigenstates, and recognize $W_L$ as the pure SM $W$-boson. 
The lower bound on the $W'$ mass comes from a variety of low-energy 
constraints, e.g. $K_L-K_S$ mass difference, $B_{d,s}-\bar{B}_{d,s}$ mixing, 
weak $C\!P$ violation etc (For a recent update on the old results, 
see Ref.~\cite{yue,masswr}). The most stringent limit on $W_R$ mass in LR 
models is for the case of same CKM mixing angles in the left and right sectors: 
$M_{W_R}>2.5$ TeV~\cite{yue}; however, this limit can be significantly lowered 
if there is no correlation between the mixing angles in the two sectors~
\cite{zetaw,morewr}. The current collider bound on $W'$ mass is around 1 TeV~\cite{pdg}. 

The neutral gauge bosons in LR model are mixtures of $W^3_{L,R}$ and $B$ 
and the mixing between the weak eigenstates of these massive 
neutral bosons is parameterized as
\begin{eqnarray}
	Z &=& \cos{\zeta_Z}Z_1+\sin{\zeta_Z}Z_2,\nonumber\\
	Z' &=& -\sin{\zeta_Z}Z_1+\cos{\zeta_Z}Z_2
\end{eqnarray}
where $Z,Z'$ are the mass eigenstates, and in the limit 
$v_L\ll \kappa,\kappa' \ll v_R$, the mixing angle is given by 
$\tan{2\zeta_Z} \simeq 2\sqrt{\cos{2\theta_W}}(M_Z/M_{Z'})^2$. Current 
experimental data constrain the mixing parameter to $<{\cal O}(10^{-4})$
and the $Z'$ mass to values $>{\cal O}({\rm TeV})$~{\cite{zetawp,masszp}. The 
current collider limit on the LR $Z'$ mass is $>998$ GeV~\cite{pdg}.  
\subsection{Mixing in the Neutrino Sector}
In the neutrino sector of LR models, due to the presence of the RH neutrinos, 
the neutrino mass eigenstates ($\nu_i, N_i$) are mixtures of the 
flavor eigenstates ($\nu_\alpha, N_\alpha$) where $i=1,2,3$ and $\alpha = e,\mu,\tau$ 
for three generations. For type I seesaw with only one additional set of 
SM singlets, this mixing can be parameterized as
\begin{eqnarray}
	\left(\begin{array}{c}
		\nu_\alpha \\
		N_\beta
	\end{array}\right) = {\cal V}_1\left(\begin{array}{c}
		\nu_i \\
		N_j
	\end{array}\right)
	\label{eq:10}
\end{eqnarray}
where ${\cal V}_1$ is a $6\times 6$ unitary matrix diagonalizing the full neutrino mass matrix in 
Eq.~(\ref{eq:type1}). Similarly, for inverse seesaw case in which we have two sets of SM 
singlet heavy neutrinos, the mixing is given by 
\begin{eqnarray}
	\left(\begin{array}{c}
		\nu_\alpha \\
		N_\beta \\
		S_\gamma
	\end{array}\right) = {\cal V}_2\left(\begin{array}{c}
		\nu_i \\
		N_j \\
		N_k
	\end{array}\right)
	\label{eq:11}
\end{eqnarray}
where ${\cal V}_2$ is a $9\times 9$ unitary matrix diagonalizing the 
neutrino mass matrix in Eq.~(\ref{eq:invseesaw}). 
Thus the weak interaction currents of light and heavy neutrinos are modified as follows:
\begin{eqnarray}
	{\cal L}_{\rm CC} &=& \frac{g}{\sqrt 2}\left
	[W^\mu_L \bar{\ell}_\alpha\gamma^\mu P_L \nu_\alpha + 
	W^\mu_R \bar{\ell}_\beta \gamma^\mu P_R N_\beta \right] + {\rm h.c.} \nonumber\\
	& = & \frac{g}{\sqrt 2}\left[W^\mu_L \bar{\ell}_\alpha 
	\gamma^\mu P_L ({\cal V}_{\alpha i}\nu_i+{\cal V}_{\alpha j}N_j)
+ W^\mu_R \bar{\ell}_\beta \gamma^\mu P_R ({\cal V}_{\beta i}\nu_i+{\cal V}_{\beta j}N_j) 
\right] + {\rm h.c.}, 
\label{eq:12} \\
{\cal L}_{\rm NC} &\simeq& \frac{g}{2\cos{\theta_W}}\left[Z_\mu \bar{\nu}_\alpha\gamma^\mu P_L \nu_\alpha 
 + \sqrt{\cos{2\theta_W}}Z'_\mu \overline{N}_\beta \gamma^\mu P_R N_\beta\right]\nonumber\\
 & = & \frac{g}{2\cos{\theta_W}}\left[Z_\mu\left\{{\cal V}^*_{\alpha i_1}{\cal V}_{\alpha i_2}\bar{\nu}_{i_1} \gamma^\mu P_L \nu_{i_2} 
 + ({\cal V}^*_{\alpha i}{\cal V}_{\alpha j}\bar{\nu}_i\gamma^\mu P_L N_j + {\rm h.c.})+ {\cal V}^*_{\alpha j_1}{\cal V}_{\alpha j_2} \overline{N}_{j_1}\gamma^\mu P_L N_{j_2}\right\}
\right.\nonumber\\
&&\left. 
+ \sqrt{\cos{2\theta_W}}Z'_\mu \left\{{\cal V}^*_{\beta j_1}{\cal V}_
{\beta j_2}\overline{N}_{j_1} \gamma^\mu P_R N_{j_2}
+ ({\cal V}_{\beta i}^*{\cal V}_{\beta j}\bar{\nu}_i\gamma^\mu P_R N_j + {\rm h.c.})+ {\cal V}^*_{\beta i_1}{\cal V}_{\beta i_2} \bar{\nu}_{i_1}\gamma^\mu P_R \nu_{i_2}\right\}
\right],\nonumber \\
&& \label{eq:13}
\end{eqnarray}
where we have dropped the subscript for ${\cal V}$ which now generically 
represents both ${\cal V}_1$ and ${\cal V}_2$ in Eqs.~(\ref{eq:10}) and 
(\ref{eq:11}) respectively. Thus, in general, ${\cal V}$ is a 
$(3+n)\times (3+n)$ unitary matrix, where $n$ stands for the number of SM 
singlets (3 for type-I and 6 for inverse seesaw). This can be decomposed into 
the following blocks:
\begin{eqnarray}
	{\cal V} = \left(\begin{array}{cc}
		U_{3\times 3} & V_{3\times n}\\
		X_{n\times 3} & Y_{n\times n}
	\end{array}\right)
	\label{eq:14}
\end{eqnarray}
where $U$ is the usual PMNS mixing matrix for the light neutrinos. 
The unitarity of ${\cal V}$ implies that 
\begin{eqnarray}
	UU^\dag + VV^\dag = U^\dag U + X^\dag X = I_{3\times 3},\nonumber\\
	XX^\dag + YY^\dag = V^\dag V + Y^\dag Y = I_{n\times n}. 
\end{eqnarray}
with $UU^\dag, Y^\dag Y \sim {\cal O}(1)$ and $VV^\dag, X^\dag X \sim 
{\cal O}(m_\nu/M_N)$. Thus in Eqs.~(\ref{eq:12}) and 
(\ref{eq:13}), the mixing between the light states, 
${\cal V}_{\alpha i}\equiv U_{\alpha i}$, and between the heavy states,  
${\cal V}_{\beta j}
\equiv Y_{\beta j}$ both are of order ${\cal O}(1)$, whereas the mixing 
between the light and heavy states, ${\cal V}_{\alpha j}\equiv 
V_{\alpha j},{\cal V}_{\beta i} \equiv X_{\beta i} \sim {\cal O}(M_D M_N^{-1})$ 
for both type I and inverse seesaw cases, which, in principle, could be large 
for TeV mass RH neutrinos and large Dirac Yukawa case. Henceforth, we will 
generically denote this mixing between light and heavy neutrinos by 
$V_{\ell N}$, and assume the other mixing elements in Eqs.~(\ref{eq:12}) 
and (\ref{eq:13}) to be ${\cal O}(1)$.   

The electroweak precision data constrain the 
mixing $V_{\ell N}$ involving a single charged lepton~\cite{lep_mix} and the current 
$90\%$ C.L. limits are summarized below:
\begin{eqnarray}
	\sum_{i=1}^3 |{V}_{eN_i}|^2 \leq 3.0\times 10^{-3}, ~ ~
	\sum_{i=1}^3 |{V}_{\mu N_i}|^2 \leq 3.2\times 10^{-3}, ~ ~
	\sum_{i=1}^3 |{V}_{\tau N_i}|^2 \leq 6.2\times 10^{-3}
\end{eqnarray}
These limits are crucial for our analysis since they determine the decay rate of the heavy neutrinos to multi-lepton 
final states, as discussed in next section. One can also get constraints on the mixing 
involving two charged leptons from lepton-flavor violating (LFV) processes~\cite{lfv_mix}
\footnote{However, these constraints can be easily evaded if, for example, each heavy neutrino mixes with a 
different charged lepton.}:
\begin{eqnarray}
\left	|\sum_{i=1}^3 {V}_{e N_i}{V}_{\mu N_i}^*\right| \leq 1.0\times 10^{-4},~ 
\left	|\sum_{i=1}^3 {V}_{e N_i}{V}_{\tau N_i}^*\right| \leq 1.0\times 10^{-2},~ 
\left	|\sum_{i=1}^3 {V}_{\mu N_i}{V}_{\tau N_i}^*\right| \leq 1.0\times 10^{-2}\nonumber
\end{eqnarray}
For the heavy neutrino mass below 100 GeV, the updated limits are summarized in Ref.~\cite{atre}. 

Another constraint for the manifest LR model comes from neutrino-less double beta decay as there 
is a new contribution involving the heavy gauge boson $W_R$ and RH Majorana neutrino~\cite{rnm}. 
For a TeV mass RH neutrino, this puts a lower bound on $M_{W_R}\geq 1.1$ TeV which  
increases as $M_N^{-1/4}$ for smaller RH neutrino mass. In this paper, we therefore mainly 
focus on a TeV mass RH neutrino. 

\section{Production and Decay of Heavy Neutrinos}
At a proton-proton collider, a single heavy neutrino can be 
produced at the parton-level, if kinematically allowed, in 
\begin{eqnarray}
	q\bar{q}' \to W^*_L/W_R \to \ell^+N (\ell^-\overline{N}),
	\label{eq:prod}
\end{eqnarray}
which has lepton-number conserving (LNC) or violating (LNV) decay modes 
depending on whether $N$ is Dirac or Majorana\footnote{In Eq.~(\ref{eq:prod}) and 
following, $\overline{N}$ should be replaced by $N$ for a Majorana RH neutrino.}.
Since $\tau$-lepton identification may be rather complicated in hadron 
colliders~\cite{cms-tau}, 
we restrict our analysis to only the light charged-leptons ($\ell=e,\mu$).  
The parton-level production cross sections, generated using {\tt CalcHEP}~
\cite{calchep} and with the {\tt CTEQ6L} parton distribution function~\cite{cteq}, are shown in Fig.~\ref{fig:1} as a function of the mass of 
$N$ for 1.5, 2 and 2.5 TeV 
$W_R$ mass (solid lines) at $\sqrt{s}=14$ TeV LHC. We also show the normalized 
production cross section $\sigma/|{V}_{\ell N}|^2$ (normalized to $| V_{\ell N}|^2=1$) 
for SM $W_L$-boson mediation (dashed line), 
which is generated only through the mixing ${V}_{\ell N}$ between the LH and RH neutrinos. 
We can clearly see that the $W_L$-mediated production is highly suppressed by the mixing; 
even for large mixing, the cross section for a heavy RH neutrino with $M_{W_R}>M_{N}\gg M_{W_L}$
is mostly dominated by the $W_R$-channel because $W_R$ 
can always decay on-shell whereas the $W$ has to be highly off-shell to produce $N$.
\begin{figure}[h!]
	\includegraphics[width=7cm]{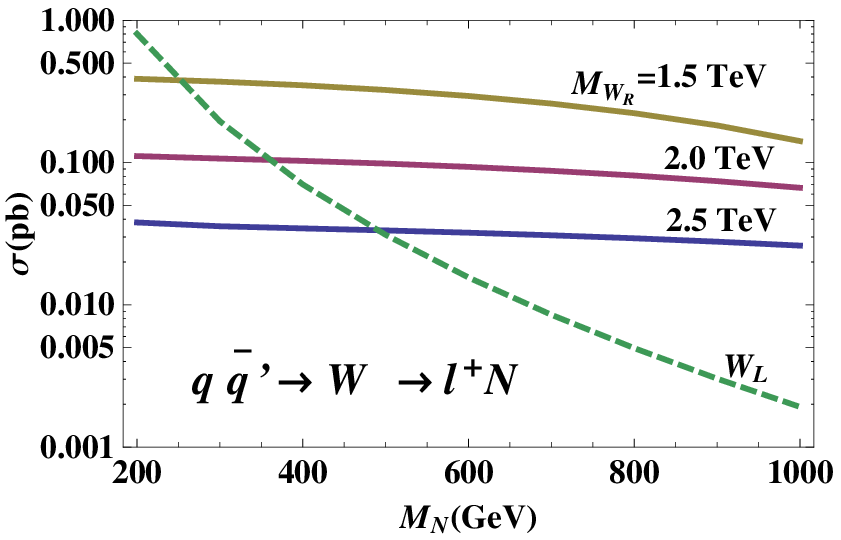}
	\hspace{1cm}
		\includegraphics[width=7cm]{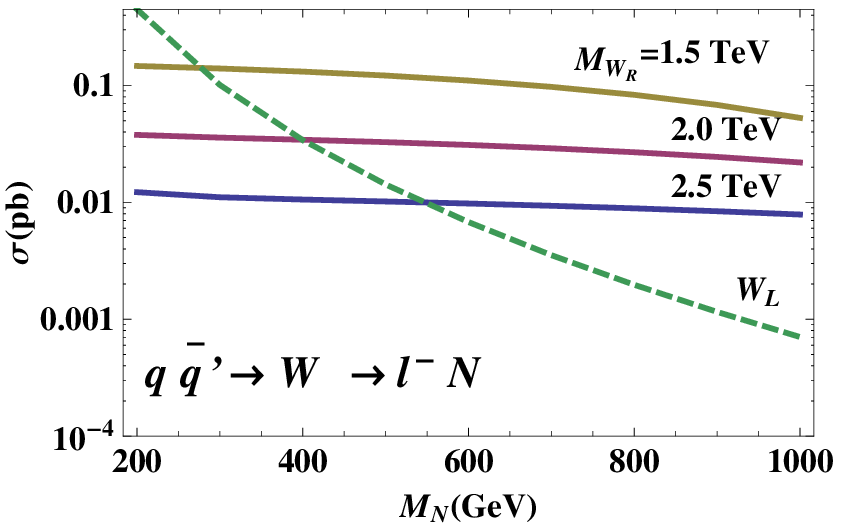}
	\caption{The cross section for $q\bar{q}' \to W^*_L/W_R \to N\ell^\pm$ for various values of $W_R$ mass (solid lines). 
	Also shown is the normalized cross section $\sigma/|{V}_{\ell N}|^2$ for $W_L$-mediated $s$-channel (dashed line).}
	\label{fig:1}
\end{figure}

The heavy RH neutrino decays to SM leptons plus a gauge or Higgs boson through its mixing with the left sector: $N\to \ell W, \nu Z, \nu H$. So all these 
decay rates are suppressed by the mixing parameter $|{V}_{\ell N}|^2$. In LR models, $N$ can also have a three-body decay mode: $N\to \ell W_R^* \to \ell jj$ 
(and similarly for $Z'$) which is not suppressed by mixing, but by mass of $W_R$. Note that the decay mode $N\to \ell W_R^* \to \ell \ell \nu$ will be 
suppressed by 
mixing as well as $W_R$-mass and hence the di-jet mode is always the dominant final state for the three-body decay of $N$. 
The various partial decay widths of $N$ are shown in Fig.~\ref{fig:2} 
for a mixing parameter $|{V}_{\ell N}|^2=0.001$ and Higgs mass of 125 GeV. 
It is clear that for mixing larger than ${\cal O}(10^{-4})$, $N$ mainly decays into the SM gauge or Higgs boson which could subsequently lead to 
multi-lepton final states.
\begin{figure}[h!]
	\includegraphics[width=8cm]{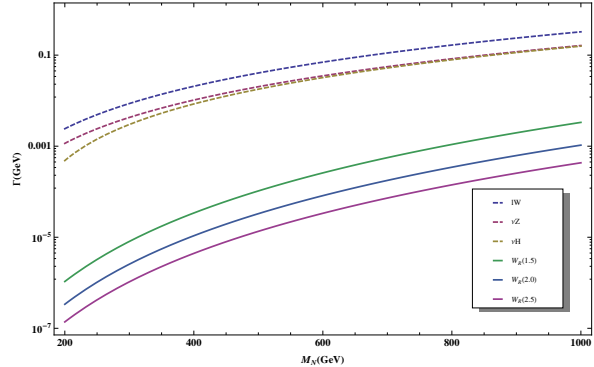}
	\caption{The partial decay widths of the RH neutrino into $\ell W, \nu Z, \nu H$ (dashed lines) as a function of its mass 
	for a mixing parameter $|{V}_{\ell N}|^2=0.001$. 
	Also shown are the three-body decay widths for $N\to \ell W_R \to \ell jj$ for $M_{W_R}=$1.5, 2.0 and 2.5 TeV.} 
	\label{fig:2}
\end{figure}

It should be emphasized here that the LR symmetry provides a unique channel for the production of RH neutrino through the $W_R$ gauge boson, without any 
mixing suppression, and multi-lepton final states through the decay of $N$ to SM gauge bosons, which even though suppressed by the mixing, still offer 
a promising channel to study the Dirac or Majorana nature of $N$. Without the LR symmetry (and hence $W_R$), the production of $N$ (through SM $W/Z$) 
will also be suppressed by mixing, which limits its observability to only a few hundred GeV masses, mainly due to the large SM background
~\cite{mtype1colb}. On the other hand, LR-symmetric models provide much higher mass reach at the LHC in the multi-lepton channel, as we discuss  
in the next section.

We further note that a single $N$ can also be produced in $q\bar q \to Z^*/Z' \to \bar{\nu} N$ but the resulting final state has either one charged 
lepton or opposite-sign di-leptons, and is buried under the huge LHC 
background~\footnote{This could, however, be important in cleaner environments, e.g. 
$e^+e^-$~\cite{almeida} and $e\gamma$~\cite{egamma} colliders.}.
One could also produce the RH neutrinos in pairs through a $Z'$-exchange: $q\bar q \to Z' \to N \overline{N}$, if kinematically allowed; however, the decay 
of two $N$'s will be suppressed by $|{V}_{\ell N}|^4$, and hence, negligible. 
\begin{figure}[h!]
	\begin{center}
\begin{tabular}{ccc}
	\includegraphics[width=5cm]{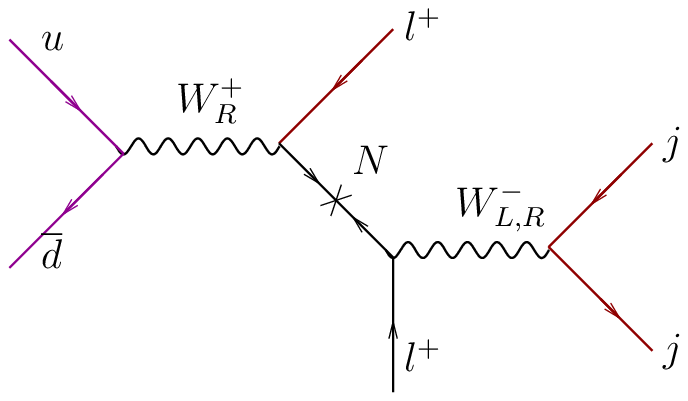} & ~ ~ ~ ~ &
	\includegraphics[width=5cm]{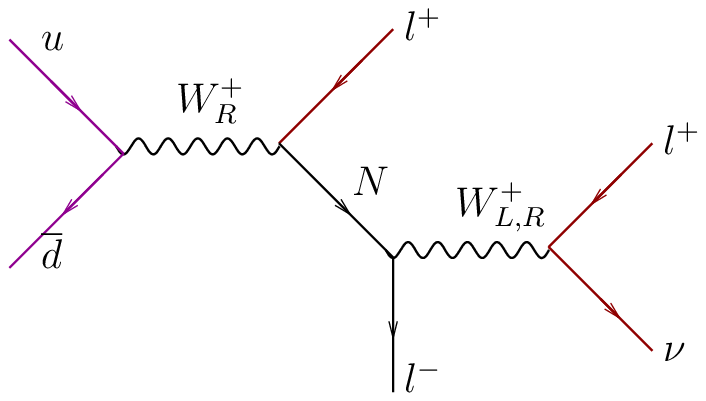} \\
	(a) Majorana $N$ & ~ ~ ~ ~ & (b) Dirac $N$
\end{tabular}
\end{center}
	\caption{The golden channels for heavy Majorana and Dirac neutrino signals at the LHC.}
	\label{fig:3}
\end{figure}

Thus we conclude from this study that for a hadron collider analysis, the most suitable production channel for a Dirac RH neutrino in LR models is 
through $W_R$-exchange and the $N$ decay mode through SM $W$. We note that this particular channel was not considered in the previous studies of RH neutrino signals in LR models~\cite{masswr, dileptonlr}, because 
they only considered a heavy Majorana neutrino (in type I seesaw) for which the golden channel is the same-sign di-lepton mode in Fig.~\ref{fig:3}(a): 
$q\bar{q}'\to W_R^{\pm} \to N\ell^{\pm} \to W^*_R\ell^{\pm}l^{\pm} \to jj\ell^{\pm}\ell^{\pm}$~\cite{keung}. In this case, the 3-body decay mode of 
$N \to \ell W_R^* \to \ell jj$ is dominant over the 2-body decay $N\to \ell W$ because the latter is suppressed by mixing which is usually very small in type I seesaw. 
However, for a heavy Dirac neutrino, this same-sign di-lepton mode is absent and the corresponding opposite-sign 
di-lepton mode $q\bar{q}'\to W_R^{\pm} \to N\ell^{\pm} \to W^*_R\ell^{\mp}l^{\pm} \to jj\ell^{\mp}\ell^{\pm}$ has large SM background. So the golden channel for a heavy Dirac 
neutrino is the tri-lepton mode in Fig.~\ref{fig:3}(b) where the $W/W_R^*$ decays to leptonic final states: 
$pp\to W_R^{\pm} \to N\ell^{\pm} \to W/W^*_R\ell^{\mp}\ell^{\pm} \to \nu \ell^{\pm} \ell^{\mp}\ell^{\pm}$~\cite{trilepton1,trilepton2}. As discussed earlier in this section, the $N$ 
decay to SM $W$ is dominant over the 3-body decay through $W_R$ for mixing $\left|{V}_{\ell N}\right|\lsim 10^{-4}$, which is easily satisfied in inverse seesaw 
models, for instance. This is also true for type I seesaw with large mixing~\cite{kersten,type13col}, in which case the 2-body decay of $N$ to SM gauge 
bosons ($W,Z,H$) will be dominant over the three-body decay through a virtual $W_R$.  

\section{Multi-Lepton Signals and SM Background}
We perform a full LHC analysis of the multi-lepton final states given in Fig.3 and the SM background associated with it. The signal and background events 
are calculated at parton-level using {\tt CalcHEP}~\cite{calchep} which are then fed into {\tt PYTHIA}~\cite{pythia} to add initial and final state 
radiation and pile up, and perform hadronization of each event. Finally, a fast detector simulation is performed using {\tt PGS}~\cite{pgs} 
to simulate a generic LHC detector. We use the more stringent {\tt L2} 
trigger~\cite{l2trig} in order to reduce the SM background. We note that the 
signal strength remains the same, if we use the low threshold {\tt L1} 
trigger, which is very close to the actual values 
used by the CMS detector. The {\tt L2} trigger has high enough thresholds to 
reduce all the SM background below the signal and therefore we do not need to 
impose any additional cuts on the events. 

The major SM background for the di-lepton signal comes from the semi-leptonic decay of a $t\bar{t}$ pair,
\begin{eqnarray}
	q\bar{q}, g\bar{g} \to t\bar{t} \to W^+b W^-\bar{b} \to jjb \ell^-\bar{\nu}\bar{b},
\end{eqnarray}
and the $b$-quark giving the second lepton: $b\to c\ell\nu$. 
Similarly, tri-lepton background is produced in the fully leptonic decay of $t\bar{t}$ and the third lepton 
coming from $b$-quark. Though the charged leptons from $b$-quark decay typically have small transverse momentum, the large $t\bar{t}$ production cross 
section (compared to the production of $N$) is responsible for the dominant background, and must be taken into account in the detector simulation. The other 
dominant SM backgrounds for multi-lepton channels at the LHC 
arise from the production of 
$WZ,WW,ZZ,WWW,Wt\bar{t},Zb\bar{b},Wb\bar{b}$ etc.. 
A detailed discussion of the 
background analysis for multi-lepton final states can be found in Ref.~\cite{trilepton1,background}. We find that by implementing the {\tt L2} trigger, most 
of this SM background can be eliminated, and the remaining 
background is dominantly due to $t\bar t,~WW,~WZ$ and $ZZ$ (which we 
collectively denote as `SM background' in the following).  

The invariant mass of the final state particles is used to reconstruct the mass of $W_R$. The selected events for the 
tri-lepton $(\ell^\pm \ell^\mp \ell^\pm)+\not{\!\!\!E_T}$ final state is shown in Fig.~\ref{fig:4} (thick lines) 
as a function of the invariant mass (100 GeV bins) for $\sqrt s = 
14$ TeV LHC and integrated luminosity, ${\cal L} = 8~ {\rm fb}^{-1}$. The expected SM background events ($t\bar{t}+VV$) are also shown (thin lines). Here we have 
chosen $M_{W_R}=2$ TeV and $M_{N} = 1$ TeV. We have also taken the mixing parameter ${V}_{\ell N}$ just below the experimental upper bound: 
$|{V}_{\ell N}|^2 = 0.0025$ (For a lower value of mixing, the cross section and hence the total number of events, will decrease as 
$|{V}_{\ell N}|^2$). We find that the invariant mass of $W_R$ is reconstructed nicely and the tri-lepton channel is virtually 
background free above 1 TeV or so. We also plot the invariant mass of 
$(\ell^\pm \ell^\mp \ell^\pm)$ in Fig.~\ref{fig:5} which has the sharp end point at $W_R$ mass. 
We note here that the tri-lepton final states with two positively charged (anti)leptons has more likelihood to be 
produced than those with one positively charged, which is naively expected for 
a proton-proton collision.
\begin{figure}[h!]
	\includegraphics[width=8cm]{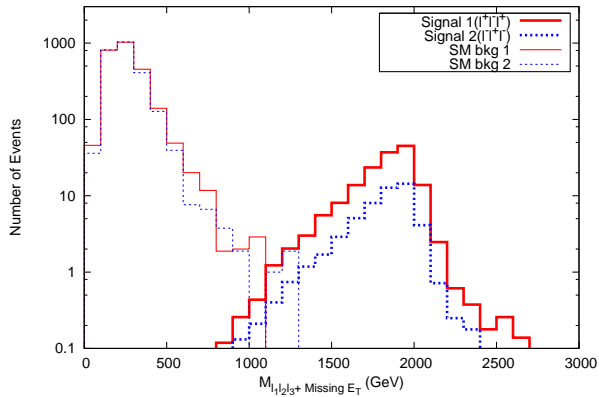}
	\caption{Selected events for the tri-lepton final state as a function of the invariant mass of  
	$\ell^\pm \ell^\mp \ell^\pm+{\not}E_T$ (100 GeV bins) for $\sqrt s = 
	14$ TeV and ${\cal L} = 8~ {\rm fb}^{-1}$. We have chosen $M_{W_R}=2, M_N=1$ TeV and $|{V}_{\ell N}|^2 = 0.0025$ for this plot. The dominant SM 
	background ($t\bar t+WW+WZ+ZZ$) is also shown here.}
	\label{fig:4}
\end{figure}
\begin{figure}[h!]
	\includegraphics[width=8cm]{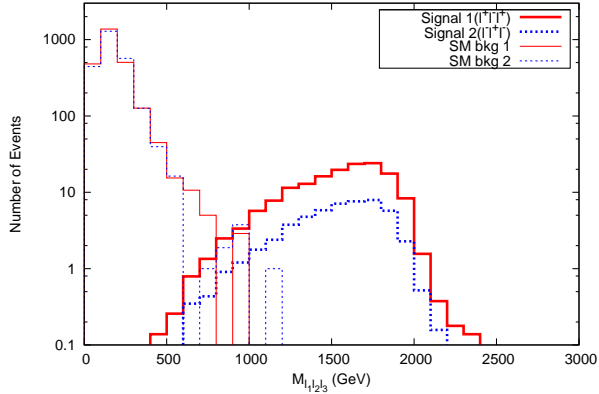}
	\caption{Selected events for tri-lepton final state as a function of the invariant mass of $\ell^\pm \ell^\mp \ell^\pm$ for the same 
	parameters as in the Fig.~\ref{fig:4} caption.}
	\label{fig:5}
\end{figure}

For comparison, we have also performed similar analysis for a heavy Majorana neutrino, similar to those in Ref.~\cite{masswr,dileptonlr}, but 
with a large mixing $|{V}_{\ell N}|^2=0.0025$. Hence, as we discussed in Sec. III, $N$ mostly decays to SM gauge bosons and charged leptons, 
and not through the 3-body decay involving $W_R$. The resulting events are shown in Figs.~\ref{fig:6} and \ref{fig:7} for the invariant mass of 
$\ell ^\pm \ell^\pm jj$ and $\ell ^\pm \ell^\pm$ respectively. The parameters chosen are the same as for Figs.~\ref{fig:4} and \ref{fig:5}. 
We note that the number of same-sign di-lepton events passing the {\tt L2} 
trigger are roughly one order of magnitude larger than the tri-lepton events. This is 
because of the overall enhancement of the cross section for the di-lepton final state because the branching fraction for hadronic decay modes of 
$W\to jj$ is roughly thrice that of the light leptonic decay modes $W\to \ell \nu$. 
\begin{figure}[h!]
	\includegraphics[width=8cm]{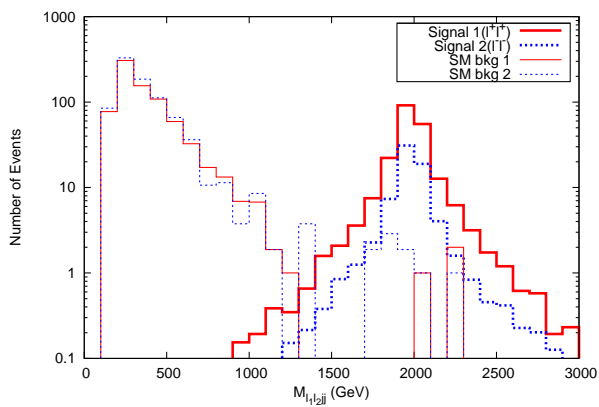}
	\caption{Selected events for the same-sign di-lepton final state as a function of the invariant mass of  
	$\ell^\pm \ell^\pm jj$ for the same parameters as in the Fig.~\ref{fig:4} caption.} 
	\label{fig:6}
\end{figure}
\begin{figure}[h!]
	\includegraphics[width=8cm]{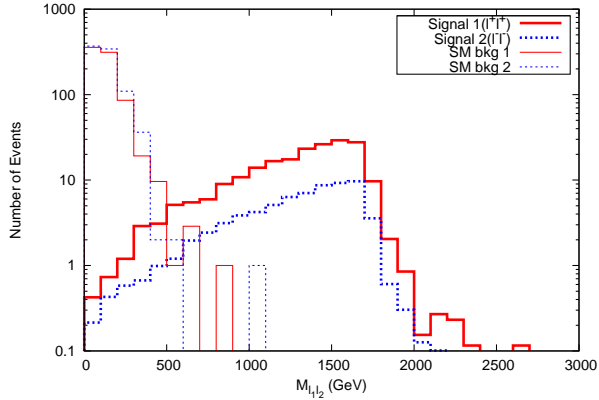}
	\caption{Selected events for same-sign di-lepton final state as a function of the invariant mass of $\ell^\pm \ell^\pm$ for the same 
	parameters as in the Fig.~\ref{fig:4} caption. }
	\label{fig:7}
\end{figure}
\section{Conclusion} 
We have discussed the collider signatures of a heavy SM singlet neutrino in a minimal LR framework, which can be of 
either Majorana or Dirac nature depending on the mechanism for neutrino mass generation. In particular, we have 
analyzed the multi-lepton signals to distinguish a TeV scale Dirac neutrino from a Majorana one at the LHC. 
We perform a detailed collider simulation to show that, in LR models, 
a TeV-scale heavy neutrino can be produced at the LHC dominantly through a $W_R$ exchange, which subsequently decays dominantly via SM gauge boson exchange. 
The invariant mass of the final state particles can be used to nicely reconstruct the mass of $W_R$ in multi-lepton channels which are 
virtually background free above a TeV. We observe that if the heavy 
neutrino is of Majorana-type, there will be distinct lepton-number violating 
signals, including the same-sign di-lepton signal discussed here. However, in 
the absence of the same-sign di-lepton signal, the tri-lepton signal can be used 
to establish the Dirac nature of the heavy neutrino. This provides a direct 
way of probing the seesaw mechanism and the associated new 
physics at TeV-scale, and can be used to distinguish type-I seesaw (with 
purely Majorana heavy neutrinos) from inverse seesaw (with pseudo-Dirac ones) 
at the LHC.

\section*{Acknowledgment} 
We would like to thank R. Sekhar Chivukula, Ayres Freitas, Tao Han, Rabindra 
Mohapatra and Lincoln Wolfenstein for very helpful discussions, 
and Fabrizio Nesti for his valuable input on detector simulation. The work of C.-Y.C. is supported by 
the George E. and Majorie S. Pake Fellowship, and the work of B.D. is partially supported by the NSF Grant No. PHY-0968854.

\end{document}